# The Double Kerr Solution as a Possible Mechanism For Controlled Global Causality Violation

Jude Prezens[1],
[1]Undergraduate Laboratories, School of Physics,
University of Melbourne, Parkville VIC 3010, Australia

Email Address: jprezens@unimelb.edu.au

**Abstract:** For over 25 years, a solution has existed to Einstein's vacuum equation that describes a space-time with two Kerr black holes. First formulated by Kramer and Neugebauer (KN) in 1980 [1], this solution has been extensively researched by many, with many and varied implications. One of which is causality violation, which I will be discussing in this paper.

## Introduction

One paper in particular is of great interest; written in 2003 by Bonner and Steadman (BS) [2]. It uses the (KN) solution to describe a region of space-time where Closed Time-like Curves (CTC) may occur. This will also be the focus of this paper. However, I intend to use a different idea to find CTC. Firstly, I will use the BS idea and corrections made by Manko and Ruiz (MR) [3], to describe the double Kerr field and to determine certain values, like the mass and angular momentum of the two black holes and also their separation. Because of the nature of the solution, the black holes can be represented as two rods separated by a larger rod (or cylinder), which also has mass and angular momentum.

Secondly, using the previously mentioned values for the central cylinder, and inputting these values into Tipler's [4] equations for a rotating cylinder, I intend to show that there is a region outside the cylinder where CTC occur, and thirdly, I propose that this is a possible mechanism for controlled causality violation.

## 1. The Double Kerr Solution

The double Kerr solution, formulated by KN in their 1980 paper [1], is a stationary axially symmetric solution to the Einstein vacuum equations. The metric is given by

$$ds^2 = -f^{-1}[e^{\nu}(dz^2 + d\rho^2) + \rho^2 d\phi^2] + f(dt - wd\phi)^2 \qquad (1)$$

where $f$, $v$ and $w$ are functions of $z$ and $\rho$ only.

This equation appears to refer to rotation, and in particular cases, they refer to two Kerr black holes (BH). NUT sources also appear, making certain solutions unphysical (for more information, refer to Appendix A of [2].) In the BS paper [2], BS also describes the occurrence of CTC near the axis of symmetry of the system, which will be discussed later in this paper.

I would like now to discuss the double Kerr solution as offered by BS. Einstein's vacuum equations for the metric (1), may be written in the Ernst formalism as

$$E = f + i\psi, \quad (2) \qquad \xi = \frac{1-E}{1+E}, \quad (3) \qquad (\xi\xi^* - 1)\nabla^2\xi = 2\xi^*(\xi_z^2 + \xi_\rho^2), \qquad (4)$$

$$w_z = \rho f^{-2}\psi_\rho, \qquad w_\rho = -\rho f^{-2}\psi_z \qquad (5)$$

where $\nabla^2$ is the Laplace operator given in cylindrical coordinates, a suffix $z$ or $\rho$ represents differentiation with respect to $z$ or $\rho$ and * represents complex conjugation. When solved for $f$ and $w$, $v$ can be determined up to a constant of integration by the left over field equations give by:

$$\rho^{-1}v_z = f^{-2}f_z f_\rho - \rho^{-2}f^2 w_z w_\rho \qquad (6)$$

$$-2\rho^{-1}v_\rho = f^{-2}(f_z^2 - f_\rho^2) + \rho^{-2}f^2(w_\rho^2 - w_z^2) \qquad (7)$$

Also, $\xi$ is the Ernst potential.

The KN solution for equation (4) seemingly refers to two Kerr BHs [1], given by:

$$\xi = \frac{N}{D} \qquad (8)$$

Where N is given by:

$$\begin{vmatrix} S_1 & S_2 & S_3 & S_4 \\ 1 & 1 & 1 & 1 \\ K_1 & K_2 & K_3 & K_4 \\ K_1^2 & K_2^2 & K_3^2 & K_4^2 \end{vmatrix} \qquad (9)$$

And D is given by:

$$\begin{vmatrix} S_1 & S_2 & S_3 & S_4 \\ 1 & 1 & 1 & 1 \\ K_1 & K_2 & K_3 & K_4 \\ K_1 S_1 & K_2 S_2 & K_3 S_3 & K_4 S_4 \end{vmatrix} \quad (10)$$

In addition:

$$S_k = e^{i\omega_k} r_k, \quad r_k = \left| [\rho^2 + (z - K_k)^2]^{1/2} \right| \quad (11)$$

$\omega_k$ and $K_k$ are real constants and k= 1, 2, 3, 4. If $K_1, K_2, K_3, K_4$ are set to zero, $\xi$ will generate a Kerr space-time. This is why this solution, (8), appears to describe two Kerr BHs.

To obtain a solution with two Kerr BHs we choose:

$$K_1 = m_1 p_1 + z_1, \quad K_2 = -m_1 p_1 + z_1 \quad (12)$$

$$K_3 = m_2 p_2 + z_2, \quad K_4 = -m_2 p_2 + z_2 \quad (13)$$

$$\omega_1 = \lambda_1, \quad \omega_2 = -\lambda_1 + \pi, \quad \omega_3 = \lambda_2, \quad \omega_4 = -\lambda_2 + \pi \quad (14)$$

$$p_s = \cos \lambda_s, \quad q_s = \sin \lambda_s, \quad -\pi/2 \leq \lambda_s \leq \pi/2 \quad (15)$$

Where s = 1, 2, $m_s$ are the BH masses and $m_s^2 q_s$ = angular momentum. If $a_s$ is the angular momentum per unit mass, then:

$$q_s = a_s / m_s \quad (16)$$

Because $q_s^2 \leq 1$ (from (15)), $m_s^2 \geq a_s^2$, meaning that hyper-extreme Kerr BHs are not included in the solution. These Kerr BHs are on the z-axis at $z = z_s$ and are represented in Weyl coordinates of (1) as two rod segments of the z-axis with lengths given by $2 m_s p_s$. (See Figure 1.)

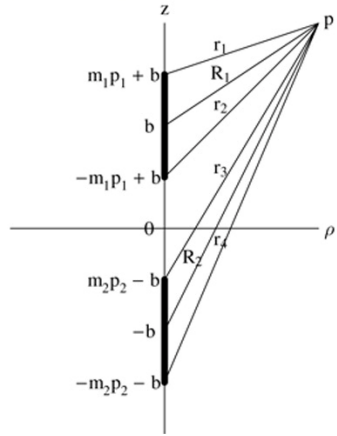

**Figure 1.**
(Taken from figure 1 of Ref. [2])

Now let
$$z_1 = -z_2 = b$$

Now we assume
$$m_1 p_1 + m_2 p_2 < 2b$$

Now we rewrite solution (8) and define real quantities A, B, P and Q by

$$\xi = \frac{N}{D} = \frac{A+iB}{P+iQ} \tag{17}$$

A, B, P and Q are given in equations (18)-(24) of [2].

In the BS paper, they make reference to another paper where infinite redshift surfaces of the double Kerr solution occur, these were said to occur when $f=0$, or when $A^2 + B^2 = P^2 + Q^2$. However, their focus was the sources, and to see whether there are regions of CTC in their neighborhood. This required BS to resort to approximations. Before doing this, they made a connection between (17) and (18)-(21) of BS [2] together with (2) and (3) to obtain:

$$E = f = \frac{b^2(r_1 + r_2 - 2m)(r_3 + r_4 - 2m) + m^2(r_1 - r_3 - 2b)(r_2 - r_4 + 2b)}{b^2(r_1 + r_2 + 2m)(r_3 + r_4 + 2m) + m^2(r_1 - r_3 + 2b)(r_2 - r_4 - 2b)} \tag{18}$$

This turned out to be $g_{44}$ for the double Schwarzschild solution i.e. for two spherical masses. (See Appendix B of [2].)

In their analysis of the solution, they introduced:

$$R_1 = \left|[\rho^2 + (z-b)^2]^{1/2}\right| \tag{19}$$

$$R_2 = \left| \left[ \rho^2 + (z+b)^2 \right]^{1/2} \right| \tag{20}$$

which refer to distances from background particles in Euclidean space (see figure 1), and they expressed the four $r_k$ as a series of terms in the two $R_s$, given by equations (28)-(31) of BS [2]. Their next step was to write A, B, P and Q, given by (18)-(21) of [2], in a power series in $m_s$, with the intention of expressing the metric functions *f, w* and *v* in a similar way. The key to their solution was to express $\xi$ as calculated from (17). This is given by:

$$\xi = \left[ \frac{m_1}{R_1} + \frac{m_2}{R_2} + O(m^3) \right] + i \left[ \frac{m_1^2 q_1 (z-b)}{R_1^3} + \frac{m_2^2 q_2 (z+b)}{R_2^3} + \frac{m_1 m_2}{b} \left( \frac{q_2}{R_1} - \frac{q_1}{R_2} \right) + O(m^4) \right] \tag{21}$$

where $O(m^n)$ means that terms of order $m_1^r m_2^p$ are neglected when $r + p \geq n$, and five parameters $(m_s, q_s, b)$ were expected. However in MR [3], this was found to be incorrect. There are actually six parameters, which I will discuss later in this paper. The imaginary part of (21) describes the sources of angular momentum and the term $m_s^2 q_s (z \mp b)/R_s^3$ refer to the spinning particles on the z-axis at $z = \pm b$ with angular momentum $m_s^2 q_s$. However, NUT sources are denoted by $\frac{m_1 m_2}{b} \left( \frac{q_2}{R_1} - \frac{q_1}{R_2} \right)$. (See Appendix A of [2].)

If there is a single NUT source, it will destroy asymptotic flatness and render the solution unphysical. However, two equal but opposite NUT sources will cancel if in (21), $q_1 = q_2$ on the z-axis except between $z = \pm b$ where they represent a finite massless spinning rod, in which case

$$q_1 = q_2 =: q \tag{22}$$

This seems to be physically significant, depending on *f, w* and *v*, though with an extra source i.e. a massless spinning rod. BS referred to this as solution I, and its Ernst potential can be written as:

$$\xi_I = \left[ \frac{m_1}{R_1} + \frac{m_2}{R_2} + O(m^3) \right] + iq \left[ \frac{m_1^2 (z-b)}{R_1^3} + \frac{m_2^2 (z+b)}{R_2^3} + \frac{m_1 m_2}{b} \left( \frac{1}{R_1} - \frac{1}{R_2} \right) + O(m^4) \right] \tag{23}$$

For the anti-parallel case where $q_1, q_2$ are equal and opposite, the NUT terms do not cancel, but they can be removed by a unitary transformation if

$$m_1 q_1 + m_2 q_2 = 0$$

or by introducing $a_s$ from (16), if

$$a_1 + a_2 = 0 \tag{24}$$

Then the Ernst potential becomes

$$\xi_{II} = \left(\frac{m_1}{R_1} + \frac{m_2}{R_2} + O(m^3)\right) + i\left(\frac{m_1^2 q_1(z-b)}{R_1^3} + \frac{m_2^2 q_2(z+b)}{R_2^3} + O(m^4)\right) \tag{25}$$

BS referred to this as solution II. However, for the purpose of this paper I will only be referring to BS solution I.

Earlier I mentioned the MR paper [3], in which they explain that in KN [1], they use "*eight real constants, $K_l$ and $\omega_l$, where l=1, 2, 3, 4, of which the constants $K_l$ can be subjected to the constraint $K_1 + K_2 + K_3 + K_4 =$ constant due to the liberty in shifting along the z-axis, while the four constants $\omega_l$ give rise to the unphysical NUT parameter which can be eliminated by the appropriate unitary transformation. The general asymptotically flat case is thus left with six arbitrary real constants (eight minus two constraints,) and not five as in [2]*", to quote MR.

A possible six-parameter axis expression for the Ernst potential given by MR [3] for an asymptotically flat double Kerr solution is:

$$\xi(\rho = 0, z) = \frac{z - b - M_1 - i(A_1 + v)}{z - b + M_1 - i(A_1 - v)} \cdot \frac{z + b - M_2 - i(A_2 - v)}{z + b + M_2 - i(A_2 + v)} \tag{26}$$

Where $M_i$ represents the Kerr particle masses, $A_i$, their angular momenta per unit mass, $i=1, 2$, $b$ is the separation constant and $v$ is related to the angular momentum part of the intermediate region.

MR goes on to explain that the BS solution does not describe a system involving super-extreme Kerr particles. The reason is because, although BS claim that they expand the KN solution only in powers of masses, they also happen to expand in powers of the angular momenta per unit mass since this too, is defined in terms of the mass parameters.

MR also says that the BS physical interpretation of the region between the two Kerr particles are separated by a massless rod (as per figure 1), but are in fact connected by a rod with mass (see figure 2) that can even intersect with the two rods that represent the two Kerr BHs. The correct choice of parameters is also important to achieve a situation that describes the KN solution.

MR then analyses the case of two Kerr BHs. If the KN solution is to have asymptotic flatness, then the so called *axis condition* needs to be the next priority. This is give by
$$\omega = 0$$

on the region $K_3 < z < K_2$ of the z-axis, the function $\omega$, entering the axisymmetric metric (1). This condition ($\omega = 0$) coverts the region $\rho = 0$, $K_3 < z < K_2$ into a massless line strut containing no CTC.

When $\omega \neq 0$ for $K_3 < z < K_2$, a system of two overlapping Kerr subextreme constituents is described (see figure 2.) In this case the region $\rho = 0$, $K_3 < z < K_2$ has, in general, a non-zero mass and angular momentum.

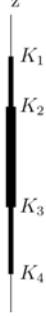

**Figure 2.**
(Taken from figure 1b of Ref. [3]).

MR then goes on to explain that the physical interpretation given by BS for solution I [2], that the *"two Kerr particles supported by a massless strut, is erroneous, because when $\omega \neq 0$ between the particles, the intermediate region $(K_3 < z < K_2)$ is not massless"*. Also, the description that the above mentioned region is *"a massless spinning rod with angular momentum $2m_1 m_2 q$ is erroneous too"*, and can be shown by calculating the Komar mass $M_K$ of any part of the rod with a formula by Akira Tomimatsu (AK) [5] in his 1983 paper, given by

$$M_K = -\frac{1}{4}\omega[\psi(z=z_1) - \psi(z=z_2)] \qquad \text{where } z_1 > z_2. \tag{27}$$

Also, $\psi$ and $\omega$ are defined by

$$\psi = -2q\left[\frac{m_1^2(z-b)}{R_1^3} + \frac{m_2^2(z+b)}{R_2^2} + \frac{m_1 m_2}{b}\left(\frac{1}{R_1} - \frac{1}{R_2}\right)\right](1-2X) + O(m^4) \tag{28}$$

$$\omega = -2\rho^2 q\left[\frac{m_1^2}{R_1^3} + \frac{m_2^2}{R_2^2}(1+X) + \frac{m_1 m_2}{R_1 R_2}\left(\frac{m_1}{R_1^2} + \frac{m_2}{R_2^2}\right)\right] + \frac{2m_1 m_2 q}{b}\left(\frac{z-b}{R_1} - \frac{z+b}{R_2}\right)$$
$$+ \frac{m_1 m_2 (m_1 + m_2) q(z^2 + \rho^2 - b^2)}{b^2 R_1 R_2} + K + O(m^4) \tag{29}$$

And X and K are defined as

$$X = \frac{m_1}{R_1} + \frac{m_2}{R_2} \tag{30}$$

$$K = -\frac{m_1 m_2 (m_1 + m_2) q}{b^2} \tag{31}$$

MR then proceeded to make a calculation of the Komar mass using arbitrary values for $m_1$, $m_2$, $b$, $q$, $z_1$ and $z_2$, with the conditions $\rho = 0$ and $|z| < b$. The value they arrived at for the Komar mass turns out to be a negative mass, which they say gives rise to the CTC in the described region. Furthermore, they describe this value as an unphysical result due to the negative mass. However, with the right selection of values for the above parameters a positive mass result can be achieved. AK also gives a formula for the Komar angular momentum give by

$$J_K = \frac{1}{4}\omega [2M_K - (2b - m_1 p_s - m_2 p_s)] \tag{32}$$

where $p_s$ is given by (15) with the condition $p_s > 0$ Using these results, it is my hypothesis that these values can be used in Tipler's equations for a rotating cylinder to achieve controlled global causality violation, which I will discuss in part 2 of this paper.

## 2. Tipler's Rotating Cylinder

In Tipler's 1974 paper [4], he defines a metric,

$$ds^2 = H(dr^2 + dz^2) + L d\varphi^2 + 2M d\varphi dt - F dt^2, \tag{33}$$

that describes the space-time around an infinitely long rotating cylinder. This space-time, with a certain configuration, gives rise to CTC. This configuration is presented below.

Again $z$ is the axis of the cylinder, $r$ is the radial distance from the axis, $\varphi$ is the angular coordinate, and $t$ must be timelike at r = 0. Also, $-\infty < z < \infty$, $0 < r < \infty$, $0 \le \varphi \le 2\pi$ and $-\infty < t < \infty$. The metric is a function of $r$ and $FL + M^2 = r^2$ is a coordinate condition that has been imposed as well as the units $G = c = 1$. $g = \det g_{\mu\nu} = -r^2 H^2$ is negative, with the metric signature being (+ + + -) for all $r > 0$, if $H \ne 0$. $\rho_T$ is the particle mass density, $a_T$ is the cylinder's angular velocity and the boundary of the cylinder is $r = R_T$, where the subscript $T$ refers to Tipler's parameters so as to avoid confusion.

For $1/2 < a_T R_T < c$

$$H = e^{-a_T^2 R_T^2} (r/R_T)^{-2a_T^2 R_T^2}$$

(34)

$$L = \frac{R_T r \sin(3\beta + \gamma)}{2\sin 2\beta \cos \beta} \tag{35}$$

$$M = \frac{r\sin(\beta+\gamma)}{\sin 2\beta} \tag{36}$$

$$F = \frac{r\sin(\beta-\gamma)}{R_T \sin\beta} \tag{37}$$

where $\beta$ and $\gamma$ are given by

$$\beta = \tan^{-1}(4a_T^2 R_T^2 - 1)^{1/2} \quad \text{and} \quad \gamma = (4a_T^2 R_T^2 - 1)^{1/2} \ln(r/R_T).$$

By using the Komar mass $M_K$, given by equation (27), we can establish the particle mass density, $\rho_T$ described by Tipler [4]. Also, using the Komar angular momentum $J_K$ given by equation (32), we can obtain the angular velocity, $a_T$ of the cylinder and thus we can determine $R_T$, the radius of the cylinder. With these values, we can calculate equations (34) – (37). According to Tipler, the CTC occur within the vicinity of the sinusoidal factors of (34) – (37). Thus, anything within the bounds of the sinusoid would move along the CTC.

Although Tipler's paper describes a cylinder that is infinitely long, he does mention that if a finite cylinder were rotating rapidly enough, CTC could be observed, and in my opinion, exploited.

## 3. Discussion

Both Kramer and Neugebauer [1] and Bonner and Steadman [2] describe a space-time containing a double Kerr black hole system. This can be described by two rods, separated by a larger central rod, or cylinder. The region around this cylinder can be described by Tipler's equations, provided the mass and angular momentum of the cylinder are first obtained in the way I have outlined in part 1 of this paper. I hypothesize that controlled global causality violation can thus be achieved.

This hypothesis could be put to the test if the LHC produces mini black holes and is able to contain them. Because these black holes would be electrically charged (positive), their mass and angular momentum could be controlled by firing electrons at them. By varying their masses, the geometry of the central cylinder would be altered slightly from one end to the other causing the resulting Tipler sinusoid to be offset. I theorise that this *offset Tipler sinusoid* will determine the direction taken along the CTC (i.e. forwards or backwards). However, because of the quantum nature of the mini black holes, this idea will need to be explored with a theory that incorporates a quantum theory of gravity, like string theory, also providing a means of testing the theory. Furthermore, *Novikov's self-consistency principle*, *Everett's many-worlds interpretation* and *Hawking's Chronology-protection conjecture* could also be tested experimentally. I hope to do this at some *time* in the future.

**Acknowledgements**

I would very much like to thank Dr. Girish Joshi for proofing my work, Archil Kobahidze for his arXiv endorsement and Sean Crosby for putting up with my questions. Thanks Girish and Dogg.